\documentstyle{article}
\parindent=0.8cm  \date{} 
\begin{document}\baselineskip 20pt
\setcounter{page}{1}
\title{\bf Variational Minimizing Parabolic Orbits for the 2-Fixed Center
Problems}
\author{{Ying Lv$^1$ and Shiqing Zhang$^2$}  \\
{\small 1.School of Mathematics and Statistics, Southwest
University, Chongqing 400715, China}\\
{\small 2.Mathematical School, Sichuan University, Chengdu 610064,
China}} \maketitle \large \maketitle \large {\bf  ABSTRACT:} Using
variational minimizing methods,we prove the existence of an odd
symmetric parabolic orbit for the 2-fixed center problems with
weak force type homogeneous potentials.

{\bf KEY WORDS:}  2-Fixed Center Problems, Odd Symmetric Parabolic Orbits,
 Variational Minimizers.

 {\bf AMS Subject Clasification:34C15,34C25.}

\section{Introduction and Main Results}
\par
Sitninkov \cite{Sitninkov} and Moser \cite{Moser} and Mathlouthi
\cite{Mathlouthi} and Souissi \cite{Souissi}and  Zhang \cite{Zhang}
etc. studied the model for the circular restricted 3-body problems:
two mass points of equal mass $m_1=m_2=\frac{1}{2}$ move in the
plane of their circular orbits such that the center of masses is at
rest, and the third small mass which does not influence the motion
of the first two ones moves on the line perpendicular to the plane
containing the first two mass points and going through the center of
mass.

Let $z(t)$ be the coordinate of the third mass point, then $z(t)$
satisfies
\begin{equation}\label{1.1}
\ddot{z}(t)+\alpha\frac{z(t)}{(|z(t)|^{2}+|r|^{2})^{\alpha/2+1}}=0.
\end{equation}
 Zhang \cite{Zhang} used variational minimizing method to prove:\\

  {\bf Theorem 1.1} \ \ \  For the equation (\ref{1.1}) with $0<\alpha<2$, there exists one
odd parabolic or hyperbolic orbit .

 The 2-fixed center problem is an old problem studied by
 Euler[6-8] etc. (\cite{Alexeev}, \cite{Macjejewski}, \cite{Marchal}, \cite{Varvoglis}):
 For two masses $1-\mu$ and $\mu$ fixed at $q^1=(-\mu,0)$
 and $q^2=(1-\mu,0)$, the problem is to study the motion $q(t)=(x(t),y(t))$
 of the third body with
 mass $m_3>0$.
 Here, we consider the motion of the third body
  attracted by the 2-fixed center masses with general homogeneous
 potentials,
 then it satisfies the following equation:
\begin{equation}\label{1.2}
\ddot{q}(t)+\frac{\partial V(q)}{\partial q}=0,
\end{equation}
\begin{equation}\label{1.3}
V(q)=-\frac{1-\mu}{|q-q^1|^{\alpha}}-\frac{\mu}{|q-q^2|^{\alpha}}.
\end{equation}

For $\mu=1/2$, we study the existence for the motion
$q(t)=(x(t),y(t))$ of the third body  satisfying
$(x(-t),y(-t))=(-x(t),-y(t)),$
here we use variational minimizing method to prove:\\

{\bf Theorem 1.2} For $(2)-(3)$ with $\mu=\frac{1}{2}$ and
$0<\alpha<2$, there exists an odd symmetrical
parabolic-type unbounded orbit.\\

\section{Truncation Functional and Its  Minimizing Critical Points }
\par

In order to find parabolic-type orbit of $(2)-(3)$ , firstly, we
restrict $t\in [-n,n]$ and find  solutions of $(2)-(3)$, then let
$n\rightarrow +\infty$ to get the parabolic-type orbit. Noticing
the symmetry of the equation, we can find the odd solutions of the
following ODE:
\begin{equation}\label{2.4}
\ddot{q}(t)=\frac{\partial U(q)}{\partial q},
\end{equation}
\begin{equation}\label{2.4}
U(q)=\frac{1/2}{|q-q^1|^{\alpha}}+\frac{1/2}{|q-q^2|^{\alpha}}.
\end{equation}
We define the functional:
\begin{equation}\label{2.4}
f(q)=\int_{-n}^{n}(\frac{1}{2}|\dot{q}(t)|^{2}+\frac{1/2}{|q-q^1|^{\alpha}}
+\frac{1/2}{|q-q^2|^{\alpha}})dt,
\end{equation}
where
\begin{equation}\label{2.4}
q\in H_{n}=\{q(t)=(x(t),y(t)):x,y\in W^{1,2}[-n,n];
 q(-t)=-q(t),q(t)\not=q^i,t\in [-n,n]\}.
\end{equation}
Since $\forall q\in H_{n}, q(0)=0$, then by the famous
Hardy-Littlewood-Polya's inequality ([9], inequality 256), for
$\forall q\in H_{n}$, we have an equivalent norm :
$$
\|q\|_{n}=(\int_{-n}^{n}|\dot{q}(t)|^{2}dt)^{1/2}.
$$\\

{\bf Remark} Here we didn't assume $q(-n)=q(n)=0$ since we want to
get the parabolic-type orbit satisfying
$$\max_{t\in R}|q(t)|=+\infty,$$
$$\min_{t\in R}|\dot{q}(t)|=0.$$
We didn't assume the periodic property for $q(t)$ since we need
non-periodic odd test function in order to get Lemma 2.6.\\

 {\bf Lemma 2.1} $f(q)$ is weakly lower semi-continuous(w.l.s.c.)
on the closure $\bar{H}_{n}$ of $H_n$.

{\bf Proof:} (i).\ It is well-known that the norm and its square
are w.l.s.c..

(ii).\ $\forall \{q_{m}\}\subset H_{n}$, if $q_{m}\rightharpoonup
q\in H_n$ weakly, then by compact embedding theorem, we have the
uniformly convergence:
$$
\max\limits_{-n\leq t\leq n } |q_{m}(t)-q(t)|\rightarrow 0,
$$
as $m\rightarrow +\infty$, so
$$
\int_{-n}^{n}\frac{1}{|q_{m}-q^i|^{\alpha}}dt\rightarrow
\int_{-n}^{n}\frac{1}{|q-q^i|^{\alpha}}dt,i=1,2,
$$
as $m\rightarrow +\infty$. Hence
$$
\mathop {\underline {\lim}}\limits_{m\rightarrow\infty}f(q_{m})\geq
f(q).
$$\\
(iii).\ $\forall \{q_{m}\}\subset H_{n}$, if $q_{m}\rightharpoonup
q\in \partial{H}_n$ weakly,let
$$ S=\{t_0\in[-n,n]:q(t_0)=q_1(t_0),or,q_2(t_0)\}$$

 (1).The Lebesgue measure of $S$ is zero,then $U(q_{m}(t)\rightarrow U(q(t))$
 almost everywhere,then by Fatou's Lemma,$\int_{-n}^nU(q)dt$ is
 $w.l.s.c.$, it is well-known that the norm and its square
are w.l.s.c.,so $f(q)$ is  $w.l.s.c.$.

(2).The Lebesgue measure of $S$:$L(S)>0$  ,then
$$\int_{-n}^nU(q)dt=+\infty,f(q)=+\infty,$$

then by compact embedding theorem, we have the uniformly
convergence on $S$:
$$
\max\limits_{-n\leq t\leq n } |q_{m}(t)-q(t)|\rightarrow 0,
$$
as $m\rightarrow +\infty$, so on $S$,we have the uniformly
convergence:

$$
\int_{-n}^{n}\frac{1}{|q_{m}-q^i|^{\alpha}}dt\rightarrow +\infty
,i=1,or,2,
$$
as $m\rightarrow +\infty$. Hence
$$
\int_{-n}^nU(q_m(t))dt\rightarrow +\infty
$$
$$
\mathop {\underline
{\lim}}\limits_{m\rightarrow\infty}f(q_{m})=+\infty\geq f(q).
$$\\

{\bf Lemma 2.2}  $f$ is coercive on $\bar{H}_{n}$.

{\bf Proof:} From the definition of $f(q)$ and Hardy-Littlewood's
inequality,it is clear that the
coercivity holds($f(q)\rightarrow +\infty,\|q\|\rightarrow +\infty $).\\

 {\bf Lemma 2.3}(Tonelli, \cite{Ambrosetti}, \cite{Mawhin}) Let $X$ be a reflexive Banach
space, $M\subset X$ be a weakly closed subset, $f:M\rightarrow
R\cup\{+\infty\}$, but $f(x)$ is not always $+\infty$ ,suppose
$f$ is weakly lower semi-continuous and coercive($f(x)\rightarrow
+\infty,\|x\|\rightarrow+\infty $),
then $f$ attains its infimum on $M$.\\

{\bf Lemma 2.4}(Palais's Symmetry Principle \cite{Palais}) Let $G$
be a finite or compact group, $\sigma$ be an orthogonal
representation of $G$, let $H$ be a real Hilbert space,
$f:H\rightarrow R$ satisfying
$$
f(\sigma\cdot x)=f(x), \forall \sigma \in G,\forall x\in H.
$$
Let
$$
F\stackrel{\triangle}{=}\{x\in H |\sigma\cdot x=x, \forall \sigma
\in G\}.
$$
Then the critical point of $f$ in $F$ is also a critical point of
$f$ in $H$.\\

{\bf Lemma 2.5} $f(q)$ attains its infimum on $ \bar{H}_{n}$, the
minimizer $\tilde{q}_{\alpha ,n}(t)$ is an odd  solution.

{\bf Proof:} Since we had proved  Lemmas 2.1-2.2, so in order to
apply for Lemma 2.3,  we need to apply for Lemma 2.4 to prove that
the critical point of $f(q)$ on $H_n$ is the odd solution of
$(4)-(5)$:
 We define groups $G_1=\{I_{2\times 2},-I_{2\times 2}\}$,$G_2=\{1,-1\}$ and their actions:

$$
\sigma_{1}\cdot q(t)=I_{2\times 2}q(t),
$$
$$
 \sigma_{2}\cdot q(t)=-I_{2\times 2}q(t);
 $$
$$
\tilde{\sigma}_{1}\cdot q(t)=q(t),
$$
$$
 \tilde{\sigma}_{2}\cdot q(t)=q(-t).
 $$

Then it's easy to prove that $f(q)$ is invariant under
$\sigma_{1},\sigma_{2},\tilde{\sigma}_{1},\tilde{\sigma}_{2},
\sigma_{i}\cdot\tilde{\sigma}_{j},\tilde{\sigma}_{j}\cdot\sigma_{i}$
and the fixed point set of the group actions for $G_1\times G_2$
 is just $H_n$, so we
can apply for Palais's Symmetrical Principle.\\

In order to get the parabolic type solution, we need to prove that
$$\tilde{q}_{\alpha,n}(t)\rightarrow\tilde{q}_{\alpha}(t)$$ when
$n\rightarrow\infty$, and $\tilde{q}_{\alpha}(t)$ has the
properties:\\
$$\max_{t\in R}|\tilde{q}_{\alpha}(t)|=+\infty,$$
$$\min_{t\in R}|\dot{\tilde{q}}_{\alpha}(t)|=0.$$
In order for that, we need some furthermore Lemmas:\\

{\bf Lemma 2.6} There exist constants $c>0$ and $0<\theta<1$
independent of $n$ such that the variational minimizing value
$a_n$ for $f(q)$ on $\bar {H_{n}}$ satisfies $a_{n}\leq
cn^{\theta}.$

 {\bf Proof:} (i). If $\tilde q(t)=(\tilde
 x,\tilde y)\in H_n$ is located on y-axis, then we choose a special odd  function
 defined by
$$
\tilde x=0,\ \tilde y=t^{\beta},\ t\in [-n,n],
$$
where $$\frac{1}{2}<\beta=\frac{l}{m}<\frac{1}{\alpha},$$ $l,m$ are
odd numbers and $(l,m)=1.$ Then
\begin{eqnarray*}
f(\tilde q(t))&=& \frac{1}{2}2\int_0^{n}\beta^{2}t^{2(\beta-1)}dt
+\int_{-n}^{n}[\frac{1/2}{|t^{2\beta}+
\frac{1}{4}|^{\alpha/2}}+\frac{1/2}{|t^{2\beta}+
\frac{1}{4}|^{\alpha/2}}]dt
\\
&\leq&\frac{\beta^{2}}{2\beta-1}n^{2\beta-1}+\frac{2}{1-\alpha\beta}n^{1-\alpha\beta}.
\end{eqnarray*}
 Now we define
\begin{equation}
\theta=\max(2\beta-1,1-\alpha\beta),\label{2.15}
\end{equation}

\begin{equation}
c=\frac{\beta^{2}}{2\beta-1}+\frac{2}{1-\alpha\beta}>0.\label{2.16}
\end{equation}
When $$\frac{1}{2}<\beta=\frac{l}{m}<\frac{1}{\alpha},$$ then
$$2\beta-1>0,\ 1-\alpha\beta>0$$ and $0<\theta<1$. Hence we have
$$f(\tilde q)\leq cn^{\theta}.$$\\
(ii). If $\tilde q(t)=(\tilde
 x,\ \tilde y)$ is not on y-axis, we choose a special odd  function on $t$ defined by
$$
\tilde{x}(t)=t^{\beta},\tilde{y}(t)=0,\ t\in [-n,n],
$$
where  $$\frac{1}{2}<\beta=\frac{l}{m}<\frac{1}{\alpha},$$ $l,m$ are
odd numbers and $(l,m)=1.$ Then, we have
\begin{eqnarray*}
f(\tilde q(t))&\leq&
\int_0^{n}\beta^{2}t^{2(\beta-1)}dt+\int_0^{n}[\frac{1}{|t^{\beta}+
\frac{1}{2}|^{\alpha}}+\frac{1}{|t^{\beta}-\frac{1}{2}|^{\alpha}}]dt
\\
&\leq&\frac{\beta^{2}}{2\beta-1}n^{2\beta-1}+[\frac{1}{1-\alpha\beta}n^{1-\alpha\beta}+
\int_0^{n}\frac{1}{|t^{\beta}-\frac{1}{2}|^{\alpha}}dt].
\end{eqnarray*}
 Now we estimate
$$
\int_0^{n}\frac{1}{|t^{\beta}-\frac{1}{2}|^{\alpha}}dt.
$$
Let
$$t^{\beta}-\frac{1}{2}=\tau^{\beta},$$\\
then $t>\tau$ and $$dt=(\frac{\tau}{t})^{\beta-1}d\tau$$also
\begin{eqnarray*}
\int_0^{n}\frac{1}{|t^{\beta}-\frac{1}{2}|^{\alpha}}dt& <&
\int_{(-\frac{1}{2})^{-\frac{1}{\beta}}}^{(n^{\beta}-\frac{1}{2})^{\frac{1}{\beta}}}{\tau}^{-\alpha\beta}d\tau\\
&<&\frac{1}{1-\alpha\beta}[n^{1-\alpha\beta}-(-\frac{1}{2})^{-\frac{1}{\beta}(1-\alpha\beta)}].
\end{eqnarray*}
Define
$$
\theta=\max\{2\beta-1,1-\alpha\beta\},
$$
$$
c=\frac{\beta^{2}}{2\beta-1}+\frac{3}{1-\alpha\beta}>0.
$$
When $$\frac{1}{2}<\beta=\frac{l}{m}<\frac{1}{\alpha},$$then
$$2\beta-1>0,1-\alpha\beta>0$$ and $ 0<\theta<1.$ Hence  we also have
$$f(\tilde q)\leq cn^{\theta}.$$
Furthermore, for our minimizer, we have

 {\bf Lemma 2.7} Let
$\tilde{q}_{\alpha,n}$ be critical points corresponding to the
minimizing critical values $a_{n}=\mathop {\min
}\limits_{H_{n}}f(q),$ then $
\|\tilde{q}_{\alpha,n}\|_{\infty}\rightarrow +\infty,$ when
$n\rightarrow +\infty$.

 {\bf Proof:}\ By the definition of
$f(\tilde{q}_{\alpha,n})$ and Lemma 2.6, we have
\begin{eqnarray*}
cn^{\theta}&\geq& f(\tilde{q}_{\alpha,n})\\
&\geq&\int_{0}^{n}[\frac{1}{|(x+\frac{1}{2})^2
+y^2|^{\alpha/2}}+\frac{1}{|(x-\frac{1}{2})^2+y^2|^{\alpha/2}}]dt.
\end{eqnarray*}
We notice that
$$(x+\frac{1}{2})^2+y^2\leq2(x^2+y^2)+\frac{5}{4},$$
$$(x-\frac{1}{2})^2+y^2\leq(x^2+y^2)+\frac{1}{4},$$
so
\begin{eqnarray*}
cn^{\theta}&\geq&\int_{0}^{n}\frac{dt}{(2\|\tilde{q}_{\alpha,n}\|_{\infty}^{2}
+\frac{5}{4})^{\alpha/2}}
+\frac{dt}{(\|\tilde{q}_{\alpha,n}\|_{\infty}^{2}+\frac{1}{4})^{\alpha/2}}\\
&\geq&\frac{2n}{(2\|\tilde{q}_{\alpha,n}\|_{\infty}^{2}+\frac{5}{4})^{\alpha/2}}.
\end{eqnarray*}
Hence
\begin{equation}\label{2.31}
\|\tilde{q}_{\alpha,n}\|_{\infty}^{2}\rightarrow+\infty,
\end{equation}
as $n\rightarrow +\infty$.\\

{\bf Lemma 2.8} $\int_a^b|\dot{\tilde{q}}_{\alpha,n}|^2dt$ is
uniformly bounded  on any compact set $[a,b]\subset R$.

{\bf Proof:} Since the system is autonomous, so for any given
$\alpha,n$, along the solution $\tilde{q}_{\alpha,n}(t)$, the
energy $h(t)$ is conservative, i.e., a constant $h=h(\alpha,n)$:

\begin{equation}\label{2.32}
\frac{1}{2}|\dot{\tilde{q}}_{\alpha,n}|^{2}-\frac{1/2}{{|\tilde{q}}_{\alpha,n}-q^1|^
{\alpha}}-\frac{1/2}{{|\tilde{q}}_{\alpha,n}-q^2|^
{\alpha}}=h.\\
\end{equation}
By the above energy relationship  and the definition of the
functional $f$, we have

\begin{eqnarray*}
 f(\tilde{q}_{\alpha,n})&=&\int_{-n}^{n}(\frac{1}{2}|\dot{\tilde{q}}_{\alpha,n}|^{2}+\frac{1/2}{|\tilde{q}_{\alpha,n}-q^1|^
{\alpha}} +\frac{1/2}{|\tilde{q}_{\alpha,n}-q^2|^{\alpha}})dt\\
&=&\int_{-n}^{n}(\frac{1}{2}|\dot{\tilde{q}}_{\alpha,n}|^{2}-\frac{1/2}{{|\tilde{q}}_{\alpha,n}-q^1|^
{\alpha}}-\frac{1/2}{{|\tilde{q}}_{\alpha,n}-q^2|^ {\alpha}})dt\\
&+&2\int_{-n}^{n}\frac{1/2}{{|\tilde{q}}_{\alpha,n}-q^1|^
{\alpha}}+\frac{1/2}{{|\tilde{q}}_{\alpha,n}-q^2|^ {\alpha}}dt\\
&=&2nh+2\int_{-n}^{n}\frac{1/2}{|\tilde{q}_{\alpha,n}-q^1|^
{\alpha}}+\frac{1/2}{|\tilde{q}_{\alpha,n}-q^2|^ {\alpha}}dt.
\end{eqnarray*}
By Lemma 2.6, we have
\begin{eqnarray*}
cn^{\theta}&\geq&
2nh+2\int_{-n}^{n}(\frac{1/2}{|\tilde{q}_{\alpha,n}-q^1|^{\alpha}}+\frac{1/2}{|\tilde{q}_{\alpha,n}-q^2|^{\alpha}})dt,\\
\end{eqnarray*}
and
\begin{equation}\label{2.37}
 h\leq\frac{c}{2}n^{\theta-1}-\frac{1}{n}\int_{-n}^{n}(\frac{1/2}{|\tilde{q}_{\alpha,n}-q^1|^{\alpha}}
 +\frac{1/2}{|\tilde{q}_{\alpha,n}-q^2|^{\alpha}})dt
\leq \frac{c}{2}n^{\theta-1}
\end{equation}
(1).When $n$ is large enough,$|\tilde{q}_{\alpha,n}(t)-q^{i}|$ has
uniformly positive lower bound,that is, $\min_{a\leq t\leq
b}|\tilde{q}_{\alpha,n}(t)-q^{i}|\geq c>0,$ then we have
\begin{eqnarray*}
\int_a^b\frac{1}{2}|\dot{\tilde{q}}_{\alpha,n}|^{2}&=&h(b-a)
+\int_a^b[\frac{1/2}{|\tilde{q}_{\alpha,n}-q^1|^{\alpha}}+
\frac{1/2}{|\tilde{q}_{\alpha,n}-q^2|^{\alpha}}]dt\\
&\leq &\frac{c}{2}(b-a)+c^{-\alpha}(b-a).
\end{eqnarray*}
(2).There exist $i_0=1$ or $2$ and a sequence ${t_n}\subset [a,b]$
such that $\tilde{q}_{\alpha,n}(t_n)\rightarrow q^{i_0}$,then
since $0<\alpha<2$,the potential is weak force potential,so when
$n$ is large,we have
$$
\int_a^b[\frac{1/2}{|\tilde{q}_{\alpha,n}-q^1|^{\alpha}}+
\frac{1/2}{|\tilde{q}_{\alpha,n}-q^2|^{\alpha}}]dt\leq M,
$$
$$
\int_a^b\frac{1}{2}|\dot{\tilde{q}}_{\alpha,n}|^{2}dt\leq
\frac{c}{2}(b-a)+M.
$$

\section{Proof of Theorem 1.2}

 By $\tilde{q}_{\alpha,n}(0)=0$ and Cauchy-Schwarz inequality and Lemma 2.8 we have
$$
|\tilde{q}_{\alpha,n}(t)|=|\int_0^t\dot{\tilde{q}}_{\alpha,n}(s)ds|
\leq
(b-a)^{1/2}[\int_a^b|\dot{\tilde{q}}_{\alpha,n}|^{2}ds]^{1/2}\leq
M_1,
$$
so we have

 (i). $\{\tilde{q}_{\alpha,n}\}$ is uniformly bounded on
any compact set of $R$. \\
By Cauchy-Schwarz inequality and Lemma 2.8 we have

$$
|\tilde{q}_{\alpha,n}(t_2)-\tilde{q}_{\alpha,n}(t_1)|=|\int_{t_1}^{t_2}
\dot{\tilde{q}}_{\alpha,n}(s)ds| \leq
[\int_a^b|\dot{\tilde{q}}_{\alpha,n}|^{2}ds]^{1/2}(t_2-t_1)^{1/2}\leq
M_2(t_2-t_1)^{1/2},
$$
so we have

 (ii). $\{\tilde{q}_{\alpha,n}\}$ is
uniformly equi-continuous on any $[a,b]\subset R$.\\
Now we can apply Ascoli-Arzel$\grave{a}$ Theorem, we know
$\{\tilde{q}_{\alpha,n}\}$ has a sub-sequence converging uniformly
to a limit $\tilde{q}_{\alpha}(t)$ on any compact set of $R$, and
$\tilde{q}_{\alpha}(t)$ is a solution of (2.2) . By the energy
conservation law and Lemmas 2.7-2.8, we have
$$
h=\frac{1}{2}|\dot{\tilde{q}}_{\alpha}|^{2}-\frac{1}{2}(\frac{1}{|\tilde
q_{\alpha}-q^1|^{\alpha}}+\frac{1}{|\tilde
q_{\alpha}-q^2|^{\alpha}}) =0.
$$
Then by Corollary 2.3 of \cite{Long}, we have

\begin{equation}\label{2.40}
\frac{1}{2}|\dot{\tilde{q}}_{\alpha}|^{2}=\frac{1/2}{|\tilde
q_{\alpha}-q^1|^{\alpha}}+\frac{1/2}{|\tilde
q_{\alpha}-q^2|^{\alpha}}
 \geq [2^{\frac{\alpha+2}{2}}][2|\tilde
q_{\alpha}|^2+\frac{1}{2}]^{-\alpha/2}.
\end{equation}
Now we claim\\
(a).
\begin{equation}\label{2.41}
max_{t\in R}|\tilde{q}_{\alpha}(t)|= +\infty.
\end{equation}
In fact, if $\exists \beta >0$ such that

$$
|\tilde{q}_{\alpha}|<\beta ,\forall t\in R.
$$
By (\ref{2.40}), $\exists \gamma >0$ such that

$$
|\dot{\tilde{q}}_{\alpha}|>\gamma, \forall t\in R.
$$
Then when $n$ is large,we have

$$
|\dot{\tilde{q}}_{\alpha,n}|>\gamma, \forall t\in R.
$$
\begin{eqnarray*}
cn^{\theta}\geq\int_{-n}^n|\dot{\tilde{q}}_{\alpha,n}|^{2}>2n\gamma^2,
\end{eqnarray*}
which is a contradiction.

Now by (\ref{2.40}) we have \\
(b).
\begin{equation}\label{2.42}
\min_{t\in R}|{\dot{\tilde{q}}_{\alpha}}(t)|= 0.
\end{equation}

\section{Acknowledgements}

 The authors sincerely thank the referee for his/her many valuable
 comments and remarks which helped us revising the paper,
  we aslo thank the supports of NSF of China and a research fund for
 the Doctoral program of higher education of China.


\begin{thebibliography}{9}
\bibitem{Sitninkov} K. Sitninkov, Existence of oscillating motion for the
three-body problem, J. Dokl. Akad. Nauk USSR 133(1960), 303-306.

\bibitem{Moser} J. Moser, Stable and random motions in dynamical
systems,Ann.Math.Studies 77, Princeton Univ. Press, 1973.

\bibitem{Mathlouthi}S. Mathlouthi , Periodic orbits of the restricted three-body
problem,Trans.AMS 350(1998), 2265-2276.

\bibitem{Souissi}C. Souissi , Existence of parabolic orbits for the
restricted three-body problem, Annals of University of Craiova,Math.
Comp. Sci. Ser. 31(2004), 85-93.

\bibitem{Zhang} S. Q. Zhang, Variational minimizing parabolic orbits
for the restricted 3-body problems, Preprint, 2010.

\bibitem{Euler} M. Euler, De motu coproris ad duo centra virium fixa
attracti. Nov. Comm. Acad. Sci. Imp. Petrop. 1766; 10: 207-42.

\bibitem{Euler1} M. Euler, De motu coproris ad duo centra virium fixa
attracti. Nov. Comm. Acad. Sci. Imp. Petrop 1767; 11: 152-84.

\bibitem{Euler2} M. Euler, Probleme un corps etant attire en raison reciproque
quarree des distances vers vers deux points fixes donnes trouver les
cas ou la courbe decrite par ce corps sera algebrique. Hist. Acad.
Roy. Sci. Bell. Lett. Berlin 1767; 2: 228-49.

\bibitem{Alexeev}  V. M. Alexeev, Generalized three-dimensional problem of
two fixed centers of gravitation-a classification of movements.
Bull. Inst. Theoret. Astron. 1965; 10: 241-71.

\bibitem{Macjejewski} A. Macjejewski, M. Przybylska,  Non-integrability of the generalized
two fixed centres problem. Celestial Mech. Dynam. Astronom. 2004;
89: 145-164.


\bibitem{Marchal} C. Marchal, On quasi-integrable problems, the example of the artificial
satellites perturbed by the Earth's zonal harmonics. Celestial Mech.
Dynam. Astronom. 1986; 38: 377-387.

\bibitem{Varvoglis} H. Varvoglis, C. H. Vozikis, K. Wodnar, The two fixed centers: An exceptional
integrable system. Celestial Mech. Dynam. Astronom. 2004; 89:
343-56.

\bibitem{Ambrosetti} A. Ambrosetti, V. Coti Zelati, Periodic solutions of
singular Lagrangian systems, Birkh$\ddot{a}$user, Basel, 1993.

\bibitem{Mawhin}J. Mawhin, M. Willem, Critical point theory and
Hamiltonian systems, Springer, 1989.

\bibitem{Palais}R. Palais, The principle of symmetric criticality, CMP
69(1979), 19-30.

\bibitem{Bolotin}S. V. Bolotin, Existence of homoclinic motions,Vestnik
Moskov Univ. ser. I Mat. Mekh. 6(1983), 98-103.

\bibitem{Chang}K. C. Chang, Infinite dimensional Morse thory and
multiple solution problems, Progress in Nonlinear Diff. Equ. and
their Appl., Vol.6, Birkh$\ddot{a}$ser, 1993.

\bibitem{Gordon} W. Gordon, A minimizing property of Keplerian orbits Amer. J.
Math. 1977; 99: 961-71.

\bibitem{Hardy} G. Hardy,  J. Littlewood, G. Polya, Inequalities. Second ed.
Cambridge Univ. Press: Cambridge 1952.

\bibitem{Long} Y. M. Long,  S. Q. Zhang, Goemetric
characterizations for variational minimization solutions of the
3--body problems, Acta Math. Sinica 16(2000), 579--592.

\bibitem{McGehee}R. McGehee, Parabolic orbits of the restricted
three-body problem, Academic Press, New York and London, 1973.

\bibitem{Poincar}H. Poincar$\acute{e}$, Les M$\acute{e}$thodes Nouvelles
de la M$\acute{e}$canique C$\acute{e}$leste, Gauthier-Villars,
Paris, 1899.

\bibitem{Rabinnowtz}P. H. Rabinnowtz, On the existence of periodic solutions for
a class of symmetric Hamiltonian systems, Nonlinear Anal. 11(1987),
595-611.

\bibitem{Rabinnowtz1}P. H. Rabinnowtz, Homoclinic orbits for
a class of Hamiltonian systems, Proc. Roy. Soc. Edinburgh Sect. A
114(1990), 33-38.

\bibitem{Sere}E. Sere, Existence of infinitely many homoclinics Hamiltonian
systems, Math. Z. 209(1992), 27-42.

\bibitem{Serra} E. Serra, S. Terracini, Collisionless periodic
solutions to some 3-body problems, Arch. Rational. Mech. Anal.
120(1992), 305-325.

\bibitem{Tanaka}K. Tanaka, Homoclinic orbits for a singular second
order Hamiltonian system, Ann. Inst. H. Poincar$\acute{e}$ Anal. Non
Lin$\acute{e}$aire 7(1990), 427-438.

\bibitem{Zhang1}S. Q. Zhang, Q. Zhou, R. Liu, New periodic solutions for
3-body problems, Celestial Mechanics and Dynamical Astronomy
88(2004), 365-378.


\end{thebibliography}
\end{document}